\documentclass[prl,superscriptaddress,twocolumn]
{revtex4}
\usepackage{graphicx} 
\newcommand{\ket}[1]{\mbox{$|#1\rangle$}}
\newcommand{\bra}[1]{\mbox{$\langle#1|$}}

\begin{document}

\title{Thermal Equilibrium as an Initial State for Quantum Computation by NMR}
\author{Amr F. Fahmy} 
\affiliation{Biological Chemistry and Molecular Pharmacology, Harvard Medical School, 240 Longwood Avenue, Boston, MA 02115, USA}

\author{Raimund Marx\footnote{Raimund Marx and Amr F. Fahmy both equally contributed to this article}}
\affiliation{Department of Chemistry, Technische Universit\"at M\"unchen, Lichtenbergstr. 4, D-85747 Garching, Germany}

\author{Wolfgang Bermel}
\affiliation{Bruker Analytik GmbH, Silberstreifen, D-76287 Rheinstetten, Germany}

\author{Steffen J. Glaser\footnote{to whom correspondence should be addressed: glaser@ch.tum.de}}

\affiliation{Department of Chemistry, Technische Universit\"at M\"unchen, Lichtenbergstr. 4, D-85747 Garching, Germany}

\begin{abstract}

\vskip 4em
We present a method of using a nuclear magnetic resonance computer to solve the Deutsch-Jozsa problem in which: (1) the number of molecules in the NMR sample is irrelevant to the number of qubits available to an NMR quantum computer, and (2) the initial state is chosen to be the state of thermal equilibrium, thereby avoiding the preparation of pseudopure states and the resulting exponential loss of signal as the number of qubits increases.  The algorithm is described along with its experimental implementation using four active qubits.  As expected, measured spectra demonstrate a clear distinction between constant and balanced functions. 
\end{abstract}

\maketitle

\section{Introduction}

Liquid state nuclear magnetic resonance (NMR) \cite{Ernst} has been used to
demonstrate basic concepts of quantum information processing, including the realization
of known quantum algorithms on a small number of qubits 
[2-7].
NMR has been leading the field both in terms of number of  qubits and in terms of coherent control techniques. Never-the-less,
liquid state NMR quantum computing is widely considered to be a futile effort because the preparation of certain initial states destroys an exponential amount of
signal \cite{Warren-PPS}  and because it is not clear how to extract useful
information by measuring expectation values. 

Here, we present concepts that make it possible to  directly use the high-temperature thermal equilibrium density operator  as an
initial state for quantum computing. We also show how to unitarily evolve this thermal state into a  state from
which useful information can be extracted.  Clearly, a key
ingredient of quantum information processing is the availability of an accessible (and
controllable) state space that grows exponentially with the number of qubits and the
ability to create superpositions of these states. This requirement applies both to pure
and mixed state quantum computing. In the case of pure states, the set of all
superpositions necessarily includes entangled states. 
However, in mixed state  quantum
ensembles, the ability to create (coherent) superpositions of an exponentially increasing
number of states does not necessarily imply  that
the density operator describing the ensemble  must be non-separable \cite{Caves} at each
point in time. In fact, we argue that the concept of separability
is of little relevance (if any) for the question whether a given
experimental implementation of a quantum algorithm is quantum or not. Furthermore,
it is important to note that for ensemble quantum computing algorithms which are based
on coherent superpositions of basis states, 
the accessible exponential
state space  is not limited by the
number of molecules in the sample.

Here, we discuss a version of the Deutsch-Jozsa algorithm 
[10-12]
for an NMR quantum computer that does not try to directly mimic a pure state
algorithm. This demonstrates that a new
set of algorithms is needed for this kind of computer. So far, there has not been an
easy way to describe algorithms or unitary evolution on the density matrix of mixed
states that is meaningful from the quantum computing point of view. Here we
introduce the concept of acting on outer products, where the use of an extra
control spin allows us to act only on one side of the outer product. This allows us to
preserve and use phase information. We show that the signal-to-noise ratio of the
presented algorithm does not diminish as the number of spins per molecule is increased,
provided the total number of molecules in the ensemble is constant. We show that this approach
provides an accessible state space that is  of exponential size in the number of
spins per molecule.
In addition to the theoretical discussion, we present the
first experimental implementation of  this version of the Deutsch-Jozsa
algorithm that does not suffer from the scaling problems associated with the preparation of pseudopure states \cite{Warren-PPS}.

\section{Computation on one vs many molecules}
By "molecule" we mean an isolated quantum system.
In the first subsection we address computation on one molecule that is restricted to be in basis states, this will correspond to classical computation. Next, we address computation still on a single molecule that is allowed to be in superposition states of the chosen basis states, this will correspond to pure state quantum computation. After this, we will talk about computation on many molecules in the two cases covered for the single molecule case. We will see that  NMR is capable of quantum parallelism over an exponential state space that is allowed to be in coherent superposition.

\subsection{Computation on a single molecule}

\subsubsection{Computation on a single molecule restricted to basis states}
In this case we have a state function that starts in some basis state (from the exponential number of computational basis states) say state $\ket{0}$. A computation consists of logical gates which correspond to transitions from the initial basis state to some other basis state 
\begin{displaymath}
\ket{\psi} = \ket{j}, \;\;0 \leq j \leq N-1
\end{displaymath}
via permutations only, where $N$ is the size of the Hilbert space. The spins are never allowed to be in a superposition state. The initial density matrix  $\ket{\psi}\bra{\psi}=\ket{j}\bra{j}$ for this system is a diagonal matrix with a single non-zero entry. During the computation, the single non-zero entry is permuted from one place to another on the diagonal. 
This corresponds to classical computation and is not more powerful than a classical computer even though a quantum system is used. 

\subsubsection{Computation on one molecule in superposition of basis states}
Here we have a system that is allowed to be in superposition of the form 
\begin{displaymath}
\vert \psi \rangle = \sum_{j=0}^{N-1} c_j \vert{j}\rangle.
\end{displaymath}
 Acting with a quantum gate (general unitary transformation) on a superposition is referred to as quantum parallelism where, in general, an exponential number of computational threads take place in parallel.  
 The density matrix  of this system is of the form $\ket{\psi}\bra{\psi}$ where non-diagonal elements are non-zero in the presence of superpositions of the basis states. The control and manipulation of superpositions is a key ingredient  that makes quantum computers more powerful than classical computers. 
 This type of computation is referred to as pure state quantum computing in the literature \cite{Deutsch1}.
    
\subsection{Computation on many molecules}
In this subsection we turn to computation using many copies of the same molecule. 
The density matrix for each case is simply defined  as the average of the  density matrices of the individual molecules in the ensemble.
 
\subsubsection{Computation on diagonal density matrices}

Here we consider  the case where  the initial state of each molecule in the ensemble is a basis state. If the initial state of each molecule happens to be the same then we have multiple copies of the case discussed above and the ensemble behaves in exactly the same way. In general however, each molecule may be in  a different initial basis state and the density matrix of the ensemble is diagonal and may have more than one non-zero diagonal entry. 
Consider an example with two molecules with one spin each with the following state functions
\begin{eqnarray*}
\ket{\psi^{(1)}} = \ket{0}, \;\; {\rm with}\; \ket{\psi^{(1)}}\bra{\psi^{(1)}} = 
\left( \begin{array}{c c}  1& 0 \\
                                     0 & 0 \\   
         \end{array}\right),     \\
\ket{\psi^{(2)}} = \ket{1}, \;\; {\rm with}\; \ket{\psi^{(2)}}\bra{\psi^{(2)}} = 
\left( \begin{array}{c c}  0& 0 \\
                                     0 & 1 \\   
         \end{array}\right).                 
\end{eqnarray*}

The corresponding density operator is 
\begin{displaymath}
\rho = 1/2
  \left(\begin{array}{c c}  1& 0 \\
                                     0 & 1\\  
         \end{array}\right).
\end{displaymath}

If the 
ensemble is operated upon by permutations which keep the density matrix diagonal, this will corresponds to parallel classical computation, where each molecule may be considered as a separate classical computer  \cite{brusch, collins1}.

Note that the number of molecules  is very important in this case since this is where the parallelism is coming from. If each molecule contains $n$ spins, the dimension of the density matrix is $2^n \times 2^n$ 
and in order for the diagonal of the density matrix  to be fully occupied, at least $2^n$ molecules will be required \cite{collins1}.  

The same conclusion holds for any ensemble that is described by a diagonal density operator throughout the computation.
A diagonal density operator does not require that
all molecules are in basis states, as off-diagonal matrix elements can average to zero. For example, for two states given by:
\begin{eqnarray*}
\ket{\psi^{(1)}} = {1 \over \sqrt{2}}(\ket{0} + \ket{1}),  {\rm with}\; \ket{\psi^{(1)}}\bra{\psi^{(1)}}  = {1 \over 2} 
\left( \begin{array}{c c}  1& 1 \\
                                     1 & 1 \\   
         \end{array}\right),    \hskip 5.5em \\
\ket{\psi^{(2)}} = {1 \over \sqrt{2}}(\ket{0} - \ket{1}),  {\rm with}\; \ket{\psi^{(2)}}\bra{\psi^{(2)}}  =  {1 \over 2}
\left( \begin{array}{c c}  \ \ 1& -1 \\
                                     -1 & \ \ 1 \\   
         \end{array}\right),         \ \ \ \  \ \ \ \ \ \ \ \ \\
\end{eqnarray*}
the density operator is also given by
\begin{displaymath}
\rho = 1/2
  \left(\begin{array}{c c}  1& 0 \\
                                     0 & 1\\  
         \end{array}\right).
\end{displaymath}

\subsubsection{Computation on off diagonal density operators}

Off-diagonal elements of the density operator can only occur if a subpopulation of the molecules exists whose state functions are superpositions of the computational basis states.
For example, in an ensemble of two molecules containing one spin each, the density operator
\begin{displaymath}
\rho = {1 \over 2}
  \left(\begin{array}{c c}  1& 1 \\
                                     1 & 1\\  
         \end{array}\right)
\end{displaymath}
requires that for each molecule, the state function has the form $
\ket{\psi^{(k)}} = {1 \over \sqrt{2}} e^{{\rm i} \theta_k} (\ket{0} + \ket{1})
$ with arbitrary phase factors $e^{{\rm i} \theta_k} $ for $k=1, 2$.

We note that an ensemble of $M$ identical molecules where 
each molecule contains $n$ spins has a density operator that is the same size as the  density operator of a single molecule  and behaves in the same manner under application of unitary transforms. If it is possible to prepare a density operator with off diagonal elements and manipulate these terms, one will have a computer as powerful as a pure state quantum computer except for the measurement step which we shall address below. In contrast to the previous example, we note that what is important for the size of the available state space is the number of spins per molecule and not the number of molecules in the ensemble.
 
\section{Ensembles of isolated quantum systems} 

In NMR implementations, 
the ensemble 
consists of  $M$ identical molecules. In every molecule, the nuclear spins
form an isolated quantum system.  Here, we assume that each
molecule contains
$n+1$ spins. (In the following sections, we will explain why we chose $n+1$ rather than
$n$ spins per molecule).
 The resonance frequency of
each spin is $\omega_l = -\gamma_l B_0$ where $\gamma_l$ is the gyromagnetic ratio of
spin $l$ and $B_o$ is the strength of the magnetic field. The state of the  spin system 
corresponding to 
molecule $k$ is given by  a wave function 
\begin{equation}
\label{equation2}
\vert \psi^{(k)} \rangle = \sum_{j=0}^{N-1} c^{(k)}_j \vert{j}\rangle
\end{equation}
where $\vert j\rangle$ are  the standard basis states used in quantum computing, $c^{(k)}_{j}$ are the corresponding amplitudes and $N = 2^{n+1}$.
The density operator of the ensemble 
is given by 
\begin{displaymath}
\rho = {1 \over{M}}\sum_{k=1}^{M} \vert \psi^{(k)} \rangle \langle \psi^{(k)} \vert.
\end{displaymath}
Each matrix element of the density operator is given by
\begin{equation}
\label{equation1}
\rho_{rs} = {1 \over M} \sum_{k=1}^{M}c^{(k)}_{r}c^{\ast(k)}_{s}.
\end{equation}

The diagonal entries of the density matrix, i.e. when  $r =s$, represent populations of the quantum 
 basis states \cite{Ernst}.  
Non-zero off-diagonal elements of the density matrix  represent {\it coherent}
superpositions of states.

If all the state functions $\vert \psi^{(k)}\rangle$  are basis states, i.e. the  quantum
system $k$ is not in a superposition  of the basis states, the product $c^{(k)}_r
c^{\ast(k)}_s$ must be zero.  Hence a necessary (albeit not sufficient) condition for an
off-diagonal element $\rho_{rs}$ to be non-zero is that  molecules  exist
whose individual quantum systems are in a superposition of the basis states $\vert {r}\rangle$ and 
$\vert {s}\rangle$, i.e. for each of these molecules the state $\vert\psi^{(k)}\rangle$ is
of the form given in equation (\ref{equation1}) with $c^{(k)}_r \neq 0$ and $c^{(k)}_s
\neq 0$. 
The ensemble is said to contain coherence \cite{Ernst} between  the basis states $\vert
r\rangle$ and $\vert s\rangle$ if the sum  of the terms $c^{(k)}_{r}c^{\ast(k)}_{s}$ 
over all molecules is non-zero, c.f. (\ref{equation1}). This implies that the
available state space is exponential in the number of spins per molecule.
One can compute with the diagonal elements only \cite {brusch}, however even in the
best case when each molecule is in a different basis state, the size of the state space
available for computation is bounded by the number of molecules in the sample \cite{collins1}.  
Indeed in this case we simply have computation by classical parallelism.
This contrasts to computing with off-diagonal elements \cite{Thermal_DJ} as we shall
see in the following.

\section{Thermal Equilibrium}

The thermal equilibrium  of an ensemble of spin systems  is described by the Boltzmann  distribution where the probability of the system being in state 
$\vert {r}\rangle$ is given by 
\begin{displaymath}
p(\vert {r}\rangle) ={ {\exp(-E_r /{\rm k}T)} \over {\sum_{j=0}^{N-1} \exp(-E_{j}/{\rm
k}T)}},
\end{displaymath}
where $E_{r}$ is the energy of the $r^{\rm th}$ eigen state of the Hamiltonian of the
system, ${\rm k}$ is Boltzmann's constant and $T$ is temperature.

If we assume that there are no coherences at thermal equilibrium \cite{Ernst}
the density operator of the system can be written as \cite{Ernst}
\begin{displaymath}
\rho_{th} \approx {{ \exp(-H/{\rm k}T)}\over{Tr(\exp(-H/{\rm k}T))}} \approx
{{1}\over{N}} ({\bf 1} - {{H}\over{{\rm k}T}})
\end{displaymath}
for $\| H \| \ll {\rm k}T$.

In a system where the size of the couplings between the spins is much less than the resonance 
frequencies $\omega_l$ of the individual spins $I_l$, the thermal density operator can be approximated
by \cite{Ernst}

\begin{equation}
\label{Eq_thermal}
\rho_{th} \approx {{1}\over{N}} ({\bf 1} - \sum_{l=1}^{n+1} \alpha_{l} I_{lz})
\end{equation}
with $\alpha_{l} = {{\hbar \omega_l} \over {{\rm k}T}}$ and 
\begin{displaymath}
I_{lz} = {{1} \over {2} }{\bf 1}  \otimes  \dots \otimes {\bf 1}  \otimes \sigma_{z}  \otimes {\bf 1}  \otimes \dots \otimes {\bf 1},
\end{displaymath}
where the Pauli matrix $\sigma_z$ appears as the $l^{th}$ term in the product.

\section{Unitary evolution, measurement and initial state preparation}

All the topics of this section are standard in the NMR literature, see e.g. \cite{Ernst}.
Application of a unitary transform $U$, whether by application of rf-pulses or by
time evolution of the coupling Hamiltonian,  to a density operator is obtained from
\begin{displaymath}
 \rho^\prime= U \rho U^{\dagger}.
\end{displaymath}
The expectation value of a Hermitian operator $A$ is
\begin{displaymath}
\langle A\rangle = Tr(A\ \rho).
\end{displaymath}
For example, for the thermal density operator $\rho_{th}$, the expectation value of
$F_z=\sum_{l=1}^{n+1} I_{lz}$ is
\begin{eqnarray*}
\langle F_{z} \rangle_{th} & = & Tr(F_{z} \ \rho_{th})\\
& = & {{1} \over {N}} \ Tr(F_{z} - F_{z} \sum_{l=1}^{n+1}{\alpha_{l}}  \ I_{1z})\\
& = &- {{{1}} \over {4N}}  \sum_{l=1}^{n+1}{\alpha_{l}}\  Tr({\bf 1}) \\
& = & -{ {1} \over {4}}  \sum_{l=1}^{n+1}{\alpha_{l}},
\end{eqnarray*}
where the facts that $F_{z}$ is traceless, $Tr({\bf 1}) = N$  and $I^2_{lx} = {{1}
\over {4}}{\bf 1}$ were used. Note that as the number $n$ of spins per molecule
increases, so will the magnitude of the measured signal $\langle F_{z} \rangle_{th}$ for $\alpha_l >0$.
In contrast, for the density operator
\begin{equation}
 \rho_{0} = {{1} \over {N}} ({\bf 1} + \alpha_{1} I_{1z}),
\end{equation}
the expectation value of $ F_{z}$ is independent of $n$:
\begin{equation}
\langle F_{z} \rangle  =  Tr(F_{z} \ \rho_0)
= { \alpha_{1} \over {4}},
\end{equation}
which is identical to $\langle I_{1z} \rangle=Tr(I_{1z} \ \rho_0)$.
(In  NMR, $\rho_0$ can be created from $\rho_{th}$ using standard
procedures, e.g.\ by a combination 
of unitary transforms and
 pulsed
field  gradients.)

Application of the Hadamard transform (or of a 90$^\circ_y$ pulse) to $\rho_{0}$ results in
the density operator
\begin{equation}
\label{equation4}
 \rho_{1} =  {{1} \over {N}} ({\bf 1} + \alpha_{1} I_{1x}),
\end{equation}
for which the expectation value of 
$ F_{x}$ (and of $I_{1x}$) is also $\alpha_1/4$.

\section{Scaling behavior}

As discussed in the previous section, for $\rho_1$ the expectation value $\langle I_{1x}
\rangle$ 
is independent of the number of spins per
molecule. Hence, the signal-to-noise ratio for a resolved resonance line of a molecule containing 3 spins is the
same as for a molecule with $10^4$ spins, assuming the same number of
molecules is in the sample, i.e. if the molar
concentration is the same.

As we saw earlier, there are no scaling problems with preparation of $\rho_1$ as an
initial state for the computation starting from the thermal density operator
$\rho_{th}$. Hence, a scalable mixed-state based quantum algorithm can be constructed if
the following two conditions hold: 

(1) Use of $\rho_1$ as an initial state for the
computation.

(2) The decision about the problem being solved is based upon the expectation
value of $I_{1x}$, where $I_{1x}$ is either parallel or orthogonal to the traceless
part of the final density operator. 

An algorithm meeting these two conditions would overcome the arguments against
scalability of NMR quantum computing \cite{Warren-PPS}. (However, just as in
pure state quantum computation, still a large number of practical or technological
impediments for realization of large-scale quantum computers would remain, such as
losses due to decoherence etc.) In the following section we will
describe such an algorithm.

\section{The Algorithm}
\label{sec3}
The starting state for the algorithm is formed by the density matrix $\rho_1$ (c.f.
Eq. (\ref{equation4})), the traceless part of which is proportional to $I_{1x}$. As indicated
in the previous sections, $\rho_1$  can be prepared  from the thermal
density operator
$\rho_{th}$ without any loss of signal as a function of the number of spins per molecule.

The key observation that allows us to  use $I_{1x}$ in the computation is
that  
$I_{1x}$ is the sum of all outer products which differ
only in the state of the first spin:
\begin{displaymath}
I_{1x} = {{1} \over {2}} \sum_{j=0}^{N'-1} |0,j\rangle \langle1,j| + |1,j\rangle \langle0,j|
\end{displaymath}
where $N' = N/2$.
Note that $j$ runs from $0$ to $N'-1$, hence the size of the state space available for calculation is $N' = 2^n$ which is 
exponential in the number of spins per molecule and is independent of the number of molecules in the sample.
We will exploit this structure and use it to apply unitary transforms to the outer products representing the states where spin 1 is in the state $|1\rangle$ only.
Application of a unitary transform $U$ (not controlled by spin 1) to $I_{1x}$ results in a sum of outer products 
where the ket and the bra do not contain information that we can directly use for the
Deutsch-Jozsa problem.
Instead we consider the effect of using controlled unitary operations $cU$.
For a unitary transform $U$ over $n$ spins, it is not hard to
construct the unitary transform $cU$ over $n+1$ spins which applies $U$ to spins
2 to $n+1$ if spin 1 is in state $|1\rangle$ and does nothing otherwise \cite{committeepaper}. The control spin
is unaffected in either case.
\begin{eqnarray*}
cU |1\rangle |j\rangle  & = & ({\bf 1} \otimes U) |1\rangle|j\rangle  = |1\rangle  U |j\rangle \\
cU |0\rangle |j\rangle & = & |0\rangle |j\rangle
\end{eqnarray*} 

For the Deutsch-Jozsa problem, given the function $f$, which is promised to be either
constant or balanced, we are given a unitary transform which acts as follows: 
\begin{displaymath}
U_f |j\rangle = (-1)^{f({j})} |{j}\rangle.
\end{displaymath} 
It is desired to find out if $f$ is balanced or constant by a single application of $U_f$.
 
The operator $cU_f$ can be used to distinguish if $f$ is a balanced or constant function
using $\rho_{1}$ as the starting state. 
There are two steps to the algorithm:
\begin{enumerate}
\item  Apply $cU_f$ to $\rho_{1}$ to obtain the density operator $\rho_{2} = {{1} \over {N}}({\bf 1} + \alpha_1\  cU_f I_{1x} cU^\dag_f)$. 
\item  Measure $I_{1x}$, i.e. find the value of $Tr(I_{1x} \rho_2)$. A value of
$\alpha_{1}/4$  will indicate that
$f$ is constant and $0$, a value of $-\alpha/4$ will indicate that the function is
constant and $1$, and a value of $0$ indicates that $f$ is a balanced function as we
shall see next.
\end{enumerate}

For the first step, apply $cU_f$ to $\rho_1$ to get
\begin{eqnarray*}
\rho_{2} &   = &  {{1} \over {N}}({\bf 1} + \alpha_1\  cU_f I_{1x} cU^\dag_f) \\ 
     &  = &  {{1} \over {N}}({\bf 1} +  {{\alpha_1} \over {2}} \sum_{j=0}^{N'-1} \{cU_f |0,j\rangle\langle1,j|cU^\dag_f   + \\ & & cU_f |1,j\rangle\langle0,j| cU^\dag_f\})\\
     &  = &  {{1} \over {N}}({\bf 1} +  {{\alpha_1} \over {2}} 
\sum_{j=0}^{N'-1} \{ |0,j\rangle\langle1,j|({\bf 1}_2 \otimes U^\dag_f)  +  \\ & &({\bf 1}_2
\otimes U_f) |1,j\rangle\langle0,j| \})\\
     &  = &  {{1} \over {N}}({\bf 1} +  {{\alpha_1} \over {2}} \sum_{j=0}^{N'-1} \{  (-1)^{f(j)} |0,j\rangle\langle1,j| + \\ & &(-1)^{f(j)} |1,j\rangle\langle0,j| \}).
\end{eqnarray*}
The second step is the measurement of  $I_{1x}$ which produces
\begin{eqnarray*}
Tr(\rho_2 I_{1x})  & = & {{\alpha_{1}} \over {4N}} \sum_{j=0}^{N'-1} \{ (-1)^{f(j)} + (-1)^{f(j)}\} \\
& = & \left \{ \begin{array}{ll}
	{{\alpha_1} \over {4}} & \mbox{if {\it f} is constant and 0} \\
	-{{\alpha_1} \over {4}} & \mbox{if {\it f} is constant and 1} \\
	0 & \mbox{if {\it f} is balanced.}
	\end{array}
	\right.
\end{eqnarray*}
As in the pure state case, the measurement produces the same answer every time it
is performed. Thus, the algorithm is deterministic. 

\section{Experiments}

Here, an experimental demonstration of the 
algorithm  is presented. The
experiments are based on the spin system of 
BOC-($^{13}$C$_2$-$^{15}$N-$^{2}$D$_2^\alpha$-glycine)-fluoride
\cite{5_Qubits} which forms a five-qubit system.  In the present application, only four of the five qubits were used explicitely
in the implemented algorithm (although it was necessary to eliminate interactions of the fifth qubit with the qubits
that play an active role in our experiments). 
For convenience, here we will use the following labels for the qubits: the nuclear
spins of the carbonyl $^{13}$C$^\prime$,
 the aliphatic
$^{13}$C$^\alpha$, the $^{19}$F, 
the $^{15}$N
and the amide $^{1}$H atoms are labeled 1, 2,
3, 4, and 5, respectively.

For our demonstration experiments, we
chose the following test functions on $n=3$ qubits: The constant test function was
\begin{equation}
f_0(x_2, x_3, x_4) = 0
\end{equation}
and the 
balanced test function was
\begin{equation}
\label{f_b}
f_b(x_2, x_3, x_4) = x_2x_3\oplus x_4.
\end{equation} 
For convenience, from now on we will drop the identity operator from the
description of the density matrix and set $\alpha_1=1$.
Starting from thermal equilibrium, the (reduced) density operator 
$\rho_0 = {I}_{1z}$ is prepared  by a sequence of INEPT-type transfers and
pulsed field gradients starting from $^1$H magnetization
(proportional to
$I_{5z}$) as explained in Ref.
\cite{5_Qubits}. Application of a $90^\circ_y$ pulse to $\rho_0$ yields the desired
initial state
$\rho_1 = {I}_{1x}$.

Practical pulse sequences that implement the 
unitary transformation $cU_{f}$ corresponding to a  given function
can be derived from an effective Hamiltonian \cite{Ernst}
\begin{equation}
\label{eff_Ham}
{\cal H}^{\it eff}={{{\rm i}}\over{\tau}} \ {\rm log} \  cU_{f}
\end{equation}
which effects the desired transformation in time $\tau$. Note that there is an infinite number of solutions
to Eq. (\ref{eff_Ham}) from which a convenient effective Hamiltonian ${\cal H}^{\it
eff}$ can be chosen that is compatible with the coupling topology of the available
spin system
\cite{Ernst}.  For the balanced test function $f_b$ defined in Eq. (\ref{f_b}),
$cU_{f_b}$ is a diagonal  unitary matrix 
in the
standard computational basis and its
 diagonal elements are $\{ 1,1,1,1,\ 1,1,1,1, \ 1,-1,1,-1,\ 1,-1,-1,1  
\}$. A convenient effective Hamiltonian corresponding to $cU_{f_b}$ is  
\begin{eqnarray*}
{\cal H}^{\it eff}_{b}={{\pi}\over{4 \tau}}  \{
{{3}\over{2}}    {\bf 1}- 3    {I}_{1z}- {I}_{2z} -
{I}_{3z} - 2{I}_{4z} 
+  2{I}_{1z}{I}_{2z} + \\ 2 {I}_{1z}{I}_{3z} + 2{I}_{2z}{I}_{3z}+ 4  {I}_{1z}{I}_{4z} -
4{I}_{1z}{I}_{2z}{I}_{3z}
\}.
\end{eqnarray*}
All product operator terms in ${\cal H}^{\it eff}_{b}$ mutually commute and hence can be implemented
independently and in arbitrary order. The term which is proportional to the identity operator can be neglected
because it only contributes an irrelevant overall phase factor to the transformation
$cU_{f_b}$.  A schematic representation of the transformations corresponding
to this decomposition of $U_{f_b}$ is shown in Fig. 1. 

\begin{figure}
\center{\includegraphics[width=3.2in,
keepaspectratio=true]{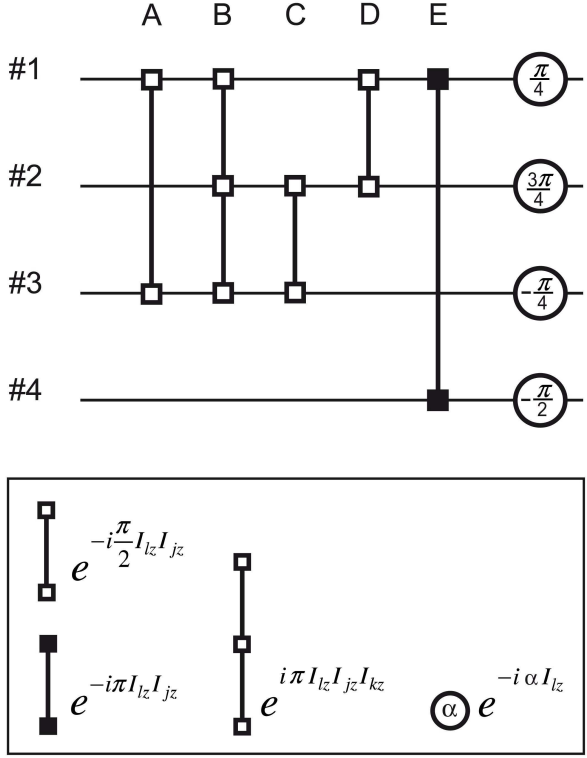}} \caption{Schematic representation of the
 decomposition of $cU_{f_b}$ given in Eq. (10). The symbols are defined in the inset.  
Up to
local spin operations (see main text for details),
the
propagators A-E correspond to the pulse sequence elements A-E of the streamlined
pulse sequence shown in Fig. 2.}
\end{figure}

The $z$ rotations
corresponding to the linear terms in  
$H^{\it eff}_{b}$ can be implemented by composite pulses 
\cite{compz} 
or by phase shifts \cite{5_Qubits}. In the
spin system of
BOC-($^{13}$C$_2$-$^{15}$N-$^{2}$D$_2^\alpha$-glycine)-fluoride,
also the terms proportional to 
${I}_{1z}{I}_{2z}$, $ {I}_{1z}{I}_{3z}$, and $ {I}_{2z}{I}_{3z}$ can be implemented in a
straight-forward way, because the required couplings
$J_{12}=
65.2$ Hz, 
$J_{13}= 
366.0$ Hz, and
$J_{23}= 
67.7$ Hz
\cite{5_Qubits} are non-zero and unwanted interactions can be removed by
decoupling. However, qubit 1 and qubit 4 are not directly coupled, which requires the
experimental simulation of the term proportional to 
${I}_{1z}{I}_{4z}$. The present implementation of this bilinear term is based on the
following  transformation of the term  ${I}_{2z}{I}_{4z}$ for which the experimental
coupling constant is $J_{24}=13.5$ Hz:
\begin{equation}
 {I}_{1z}{I}_{4z} = V ({I}_{2z}{I}_{4z}) V^{-1}
\end{equation}
with 
\begin{eqnarray*}
 V={\rm exp}\{  - {\rm i} {{\pi}\over {2}} I_{1x} \} 
 {\rm exp}\{  - {\rm i} \pi I_{1z} I_{2z} \} 
 {\rm exp}\{  - {\rm i} {{\pi}\over {2}} (I_{1y} +I_{2x})  \} \\
 {\rm exp}\{  - {\rm i} \pi I_{1z} I_{2z} \} 
{\rm exp}\{  - {\rm i} {{\pi}\over {2}} I_{2y} \} .
\end{eqnarray*}
Finally, the trilinear term in the effective Hamiltonian which is
proportional to ${I}_{1z}{I}_{2z}{I}_{3z}$ was synthesized based on the 
transformation \cite{Havel_tril}
\begin{equation}
2 {I}_{1z}{I}_{2z}{I}_{3z} = W ({I}_{1z}{I}_{2z}) W^{-1}
\end{equation}
with 
\begin{equation}
 W={\rm exp}\{  - {\rm i} {{\pi}\over {2}} I_{1x} \} \ 
 {\rm exp}\{  - {\rm i} \pi I_{1z} I_{3z} \} \ 
{\rm exp}\{  - {\rm i} {{\pi}\over {2}} I_{1y} \} .
\end{equation}

The resulting pulse sequence elements can be simplified by standard procedures \cite{5_Qubits}, 
e.g. by  eliminating adjacent $180^\circ$
pulses of the same phase or by transforming some pulses into $z$ rotations that can be implemented
by phase adjustments.
For example,
$z$ rotations (by angle $\varphi$) can be implemented by a corresponding
negative rotation
of the respective rotating frame of reference. In practice, this
results in an additional phase shift (by angle $- \varphi$) of all
following r.f.\ pulses that are applied to this spin and of the
receiver phase of this spin \cite{5_Qubits}. 
The theoretically derived pulse sequence elements for the individual terms in ${\cal H}^{\it eff}_{b}$
were thoroughly tested experimentally on numerous initial density operator terms.
A streamlined version of the pulse sequence implementing $cU_{f_b}$ is
shown in Fig. 2.

The durations of the pulse sequences for each term in the effective Hamiltonian were chosen
to be integer multiples of $\Delta=1/\vert \nu_1-\nu_2\vert=81.75\
\mu$s, in order to align the rotating frames of the homonuclear $^{13}$C
spins corresponding to qubits 1 and 2. The same pulse amplitudes and
shapes were used for hard and selective pulses as in Ref. \cite{5_Qubits}. The
bell-shaped symbols labeled ``e'' and ``g'' represent selective e-SNOB 90$^\circ$ pulses
\protect\cite{esnob}  (not for the usual 270$^\circ$, but for a $90^\circ$
rotation with a duration of 224
$\mu$s)
and
selective Gaussian 180$^\circ$ pulses \protect\cite{gauss}  (with a duration of 250
$\mu$s and a truncation level of 20\%),
respectively \cite{5_Qubits}. In order to avoid nonresonant
effects  of these selective $^{13}$C pulses, a standard compensation scheme was used
\cite{McCoy}.  For example, during an e-SNOB 90$^\circ$ pulse applied to qubit 1 (offset
0 Hz), a second e-SNOB 90$^\circ$ pulse was applied at an offset frequency of -24.462
kHz, which cancels the nonresonant effects at the offset frequency of qubit  2 at the
offset -12.231 kHz. During the selective e-SNOB 90$^\circ$ pulses, qubit 3 was
actively decoupled by applying an MLEV-4 \protect\cite{dec1,dec2,dec3} 
expanded sequence of 180$^\circ$ pulses to spin 3 (boxes labeled
``M'').  During the entire  experiment deuterium decoupling was
applied using a WALTZ-16 sequence with $\nu_{rf}=0.5$ kHz.

\begin{figure}
\center{\includegraphics[width=3.4in,
keepaspectratio=true]{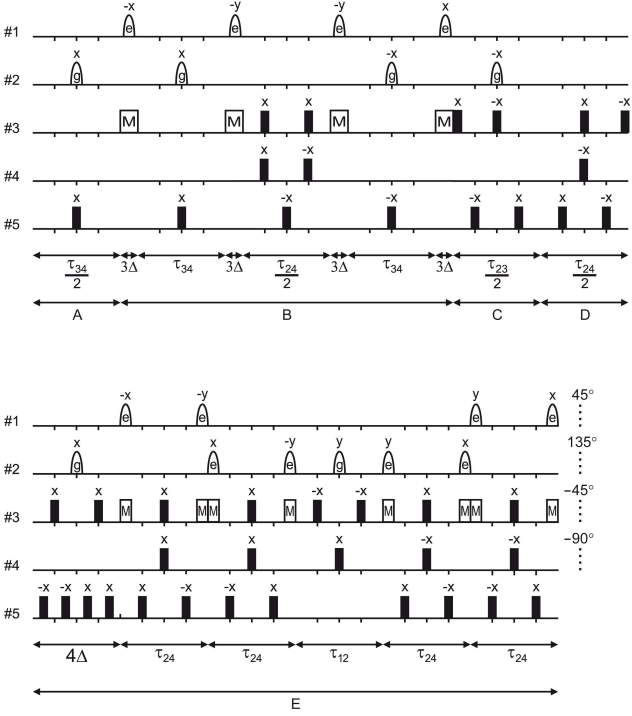}} \caption{Streamlined pulse sequence for the implementation of the unitary
transformation
$cU_{f_b}$.  The durations $\tau_{kl}=1/(2 J_{kl})$ and  $\tau_{kl}/2=1/(4 J_{kl})$  were
chosen to be integer multiples of $\Delta=1/\vert
\nu_1-\nu_2\vert=81.75\ \mu$s \cite{5_Qubits}.
Filled rectangles represent 180$^\circ$ pulses.
The bell-shaped symbols labeled ``e'' and ``g''
represent selective e-SNOB 90$^\circ$ pulses \protect\cite{esnob} and
selective Gaussian 180$^\circ$ pulses \protect\cite{gauss},
respectively. During the selective e-SNOB 90$^\circ$ pulses, qubit 3 was
actively decoupled by applying an MLEV-4 \protect\cite{dec1,dec2,dec3}
expanded sequence of 180$^\circ$ pulses to qubit 3 (boxes labeled
``M''). Dotted vertical lines represent local  $z$ rotations.
The overall duration of the pulse sequence is 84 ms.}
\end{figure}

Segments A, B,
C, D, and E  are related to the bilinear or trilinear terms of the effective Hamiltonian ${\cal
H}^{\it eff}_{b}$. Here, all propagators are implemented up to an overall (and
irrelevant) phase factor. The propagator exp$\{- {\rm i} {{\pi}\over{2}} I_{1z} I_{3z}  \}$
is implemented by segment A and additional $(\pi)_{2x}$ and $(\pi)_{5x}$ rotations
(shorthand notation for 180$_{x}^\circ$ rotations applied to qubits 2 and  5) at the end
of the segment. The propagator exp$\{{\rm i} \pi I_{1z} I_{2z} I_{3z}  \}$ is
implemented by segment B and additional $(\pi)_{2x}$ and $(\pi)_{5x}$ rotations at the
beginning of the segment and
$(\pi)_{2x}$ and $(\pi)_{1z}$ rotations at the end of the segment. 
The propagator exp$\{- {\rm i} {{\pi}\over{2}} I_{2z} I_{3z}  \}$ is implemented by
segment C and an additional $(\pi)_{2x}$ rotations at the beginning of the segment.
The propagator exp$\{- {\rm i} {{\pi}\over{2}} I_{1z} I_{2z}  \}$ is implemented by
segment D and an additional $(\pi)_{4x}$ rotations at the end of the segment.
The propagator exp$\{- {\rm i} {\pi} I_{1z} I_{4z}  \}$ is implemented by
segment E and an additional $(\pi)_{4x}$ rotation at the beginning of the segment and
a $(\pi)_{2z}$ rotations at the end of the segment. 
The additional $(\pi)_{2x}$ and $(\pi)_{5x}$ rotations at the end of
segment A and at the beginning of segment B cancel and have been eliminated in the streamlined pulse sequence. Similarly, the
remaining additional $(\pi)_{1x}$ and $(\pi)_{2x}$ rotations cancel.
The additional $(\pi)_{1z}$ and $(\pi)_{2z}$ rotations commute with the remaining
propagators and can be shifted to the end of the sequence, where they have been
combined with the linear terms of ${\cal H}^{\it eff }_{b}$.

Application of this pulse sequence to $\rho_1=I_{1x}$ yields 
\begin{eqnarray*}
\rho_1=I_{1x} \ \ \stackrel{cU_{f_b}}{\longrightarrow} \ \ \rho_b^\prime=   
{I}_{1x}{I}_{4z}+2{I}_{1x}{I}_{2z}{I}_{4z}+
2{I}_{1x}{I}_{3z}{I}_{4z}- \\ 4{I}_{1x}{I}_{2z}{I}_{3z}{I}_{4z}.
\end{eqnarray*}
None of the product operator terms in $\rho_b^\prime$ gives rise to detectable in-phase signal, i.e. the
expectation value
$\langle I_{1x}\rangle = {\rm Tr}\{
\rho_b^\prime  I_{1x}\}$ vanishes.

\begin{figure}
\center{\includegraphics[width=3.5in,
keepaspectratio=true]{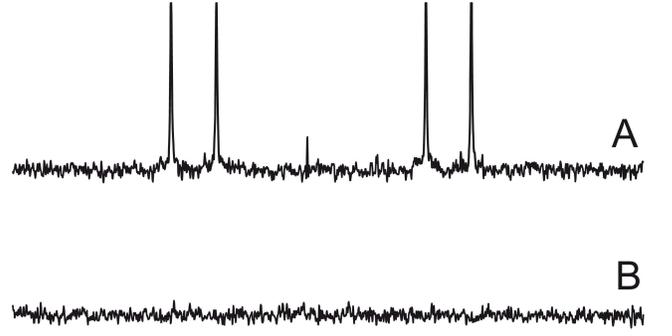}} \caption{Experimental results for (A)  the constant function $f_0$ and (B) the
balanced function $f_b$. The spectra show the signal of the carbonyl
$^{13}$C$^\prime$ spin of
BOC-($^{13}$C$_2$-$^{15}$N-$^{2}$D$_2^\alpha$-glycine)-fluoride \cite{5_Qubits}
corresponding to qubit 1.  Our NMR experiments were performed using a BRUKER AVANCE
400 spectrometer with five independent r.f.\ channels and a QXI probe (H,C-F,N). The
lock coil was also used for deuterium decoupling utilizing a lock switch. 
The spectra with a spectral
width of 900 Hz were acquired (128 scans) after applying the corresponding unitary transformations
$cU_{f_0}$ and
$cU_{f_b}$ to
$\rho_1 = {I}_{1x}$. }
\end{figure}

In general, a single time measurement of $\langle I_{1x}\rangle$ is sufficient to
distinguish balanced from constant functions because $\langle I_{1x}\rangle$ is
proportional to the integrated signal of the detected qubit in the frequency domain. If a
full free induction decay (FID) is detected, it is possible to decouple all other qubits from
the detected qubit 1, which makes the signal intensity of the detected spin (after Fourier
transformation)  independent of the number of coupled spins in the molecule. In our
present demonstration experiments, decoupling was not used during the detection of
the FID
 in order to  show the details of the full multiplet structure of the signal of the detected 
qubit 1. 
In the special case of the coupling topology of
BOC-($^{13}$C$_2$-$^{15}$N-$^{2}$D$_2^\alpha$-glycine)-fluoride, the operator $\rho_b^\prime$ cannot evolve into any
observable signal under the free evolution operator because qubits 1 and 4 are not directly coupled. Hence, even if a full FID is acquired
(rather than a single point measurement of $\langle I_{1x}\rangle$), no signal is
expected.
 The experimental spectrum shown in Fig.\ 3B matches these
theoretical predictions. For the balanced function, the spectrum in Fig. 3A shows
four resolved inphase signals (a doublet of doublets due to the couplings $J_{12}$
and $J_{13}$) with the expected intensity ratio of  1:1:1:1.
\smallskip

\begin{figure}
\center{\includegraphics[width=3.6in,
keepaspectratio=true]{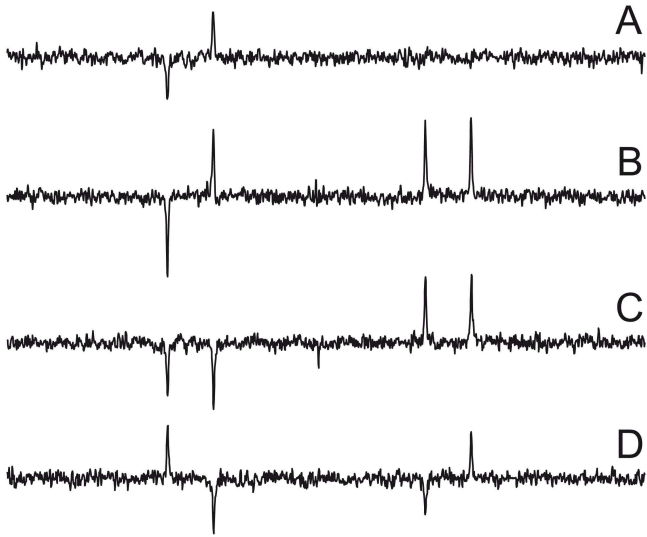}} \caption{Set of control spectra (128 scans) of qubit 1 after
application of  a ${\it CNOT}_{42}$ gate (with a duration of 37 ms) to 
 $\rho_b^\prime$ (A)  and
application of the pulse sequence of $cU_{f_b}$ (c.f. Fig. 2) to $
2I_{1x}I_{4z}$  (B).
 The additional control spectra result if the pulse sequence corresponding to
$cU_{f_b}$ is applied to
 $\rho_0=2 I_{1z}I_{3z}$ (C) and to
 $\rho_0=4 I_{1z}  I_{2z} I_{3z}$ (D), 
followed by a $90^\circ_y$ read out pulse.}
\end{figure}

In order to demonstrate that the lack of signal in Fig. 3B is not simply due to artefacts
or e.g.  relaxation losses during the pulse sequence that implements
$cU_{f_b}$ (c.f. Fig. 2), we performed a series of control experiments.
Application of the controlled-not operation ${\it CNOT}_{42}$ to $\rho_b^\prime$ yields
\begin{equation}
\rho_b^\prime \ \ \stackrel{{\it CNOT}_{42}}{\longrightarrow} \ \ \rho_b^{\prime
\prime}=  {I}_{1x}{I}_{4z}+ \underline{ {I}_{1x}{I}_{2z}}+
2{I}_{1x}{I}_{3z}{I}_{4z}- 2\underline{{I}_{1x}{I}_{2z}{I}_{3z}},
\end{equation}
where the underlined terms  give rise to a doublet of doublet with  an expected intensity ratio of $-1:1:0:0$.
The corresponding experimental  spectrum is shown in Fig. 4A. 

If $cU_{f_b}$ is applied to the inititial density operator term $2I_{1x} I_{4z}$
(rather than to $\rho_1=I_{1x}$), we find
\begin{equation}
2I_{1x} I_{4z} \ \ \stackrel{cU_{f_b}}{\longrightarrow} \ \  
{{1}\over{2}} \underline{{I}_{1x}}
+\underline{{I}_{1x}{I}_{2z}}
+\underline{{I}_{1x}{I}_{3z}}
-2 \underline{{I}_{1x}{I}_{2z}{I}_{3z}},
\end{equation}
where the underlined terms  give rise to a doublet of doublet with  an expected intensity ratio of $-1:1:1:1$. The corresponding
experimental spectrum is shown in Fig. 4B. As  ${\cal H}^{\it eff}_{b}$ contains 
only longitudinal operators, it commutes with all density
operator terms that contain only longitudinal operators. Hence, such
density operator terms are invariant under the action of
$U_{f_b}$. For example, for an initial density operator
$2 I_{1z} I_{3z} $ we expect
\begin{equation}
2 I_{1z} I_{3z}  \ \ \stackrel{cU_{f_b}}{\longrightarrow} \ \ 2 I_{1z} I_{3z} \ \  
\stackrel{90^\circ_{y} (1)}{\longrightarrow} \ \ 2\underline{I_{1x}{I}_{3z}}
\end{equation}
with  an expected  $-1:-1:1:1$ multiplet (c.f. Fig. 4C) and for an initial density operator
$4 I_{1z} I_{2z} I_{3z}  $ we expect
\begin{equation}
4 I_{1z} I_{2z} I_{3z}  \ \ \stackrel{cU_{f_b}}{\longrightarrow} \ \ 4 I_{1z}
I_{2z} I_{3z}  \ \  \stackrel{90^\circ_{y} (1)}{\longrightarrow} \ \ 4\underline{I_{1x}
I_{2z} I_{3z} }
\end{equation}
with  an expected  $1:-1:-1:1$ multiplet (c.f. Fig. 4D). In all cases, a reasonable match is found between experimental results and
theoretical predictions.

\vskip 2cm
\section{Conclusion}

We introduced the concept of acting on outer products where the use of a control
spin allows us to act only on one side of the outer product. This provides a lucid
approach to construct and analyze a version of the Deutsch-Jozsa algorithm that is
scalable in an NMR setting, where  the quantum ensemble is initially at thermal
equilibrium in the high temperature limit. The accessible state space is
of exponential size in the number of spins per molecule and the available state
space is not limited by the number of molecules in the ensemble.
Most importantly, there is no exponential signal loss with the number of qubits.

The resulting algorithm has
been implemented experimentally and a good match was found between theory and
experiment. 
A crucial point in the implementation of the
unitary transformation $cU_{f_b}$ was the simulation of trilinear and bilinear coupling
terms. The time-optimal simulation of bilinear \cite {geodesics} and  trilinear \cite{time-opt-tril} 
coupling terms can be
used to further reduce the duration of pulse sequence elements used to implement the
algorithm. The size of experimentally accessible liquid-state NMR quantum
computers is not limited by fundamental scaling problems related to the preparation of
an initial state, but by practical obstacles, such as the chemical synthesis of suitable
molecules, the control of experimental imperfections and losses through decoherence and
dissipation. However, similar questions are faced by all approaches to experimental
quantum information processing.
The fact that the Deutsch-Jozsa problem can be decided by a probabilistic
classical algorithm should not be taken as a point against NMR. Since NMR
technology can be used to decide on this problem deterministically by  a single evaluation
of the function, this should be an encouraging point to further study NMR and its capacity
to solve other problems.
It is an interesting open question if in general the thermal state can be used as an initial
state for other algorithms and whether measuring expectation values of spin operators
is useful for solving other problems.

\vskip 2em
\section{Acknowledgments}
We thank J. Myers for numerous critical discussions. A. F. thanks Gerhard Wagner
for support. S.J.G. thanks the integrated EU programme QAP, the DFG (Gl 203/4-2) and the Fonds der Chemischen Industrie for support.


\begin{thebibliography}{[28]}



\bibitem{Ernst} R.~R. Ernst, G. Bodenhausen, and A. Wokaun, 
{\em Principles of Nuclear Magnetic Resonance in One and Two
Dimensions} (Oxford University Press, Oxford, 1987).


\bibitem{gradientPPS1} D.~G. Cory, A.~F. Fahmy, and T.~F. Havel,  {\it Proc.\ of the 4th Workshop on
Physics and Computation} (New England Complex Systems Institute,
Boston, MA, 1996), pp.~87--91.

\bibitem{gradientPPS2} D.~G. Cory, A.~F. Fahmy, and T.~F.
Havel, Proc.\ Natl.\ Acad.\ Sci.\ USA {\bf 94}, 1634 (1997).

\bibitem{logicalPPS} N.~A. Gershenfeld, and I.~L. Chuang, Science
{\bf 275}, 350 (1997).


 \bibitem{Jones-Rev3}
J. A. Jones, Phys. Chem. Comm. 11,  11 (2001). 
%

\bibitem{SG-Rev}
S. Glaser, 
Angew. Chem. Int. Ed. 40, 147 
(2001).

   \bibitem{NMR-Qc-Rev1}
C. Ramanathan, N. Boulant, Z. Chen, D. G. Cory, I. L. Chuang, M. Steffen,
Quant. Inf. Proc. 3, 15 
(2004).


\bibitem{Warren-PPS}
W. S. Warren, Science {\bf 277}, 1688 (1997).


\bibitem{Caves}
S.~L.~Braunstein, C.~M.~Caves, R.~Jozsa, N.~Linden, S.~Popescu and R.~Schack,
  Phys.\ Rev.\ Lett.\  {\bf 83}, 1054 (1999).


\bibitem{deutsch} D. Deutsch, and R. Jozsa, Proc.\ Roy.\ Soc.\ London A
{\bf 439}, 553 (1992).

\bibitem{cleve} R. Cleve, A. Ekert, C. Macchiavello, and M. Mosca,
Proc.\ Roy.\ Soc.\ London A {\bf 454}, 339 (1998).

\bibitem{collins} D. Collins, K.~W. Kim, and W.~C. Holton,
Phys.\ Rev.\ A {\bf 58}, R1633 (1998).

\bibitem{Deutsch1}
D. Deutsch, Proc. R. Soc. Lond. A {\bf 400}, 97 (1985).

\bibitem{brusch}
M. Woodward, R. BrŸschweiler, quant-ph/0006024 (2000).


\bibitem{collins1}
Arvind, David Collins, Phys. Rev. A {\bf 68}, 052301 (2003). 


\bibitem{Thermal_DJ} 
 J.~M. Myers, A.~F. Fahmy, S.~J. Glaser and R. Marx, 
 Phys.\ Rev. A {\bf  63}, 032302 (2001).


\bibitem{committeepaper}
A. Barenco, C.H. Bennett, R. Cleve, D. DiVincenzo, N. Margolus, P. Shor, T. Sleator, J. Smolin, H. Weinfurter, Phys. Rev. A {\bf 52}, 3457 (1995).

\bibitem{5_Qubits}
 R. Marx, A.~F. Fahmy, J.~M. Myers, W. Bermel and S.~J. Glaser, 
Phys.\ Rev. A {\bf 62}, 012310 (2000). 


\bibitem{compz} R. Freeman, T.~A. Frenkiel, and M. Levitt, J. Magn.\
Reson.\ {\bf 44}, 409 (1981).

\bibitem{Havel_tril}
C. H. Tseng, S. Somaroo, Y. Sharf, E. Knill, R. Laflamme, T. F. 
Havel, D. G.  Cory, Phys. Rev.
A {\bf 61}, 012302 (2000).


\bibitem{esnob} ${\overline {\rm E}}$. Kup\v ce, J. Boyd, and
I.~D. Campbell, J. Magn.\ Reson.\ B {\bf 106}, 300 (1995).

\bibitem{gauss} C. Bauer, R. Freeman, T. Frenkiel, J. Keeler, and
A.~J. Shaka, J. Magn.\ Reson.\ {\bf 58}, 442 (1984).

\bibitem{McCoy} M.~A. McCoy, and L. M\"uller, J. Magn.\ Reson.\ {\bf
99}, 18 (1992).


\bibitem{dec1} M.~H. Levitt, R. Freeman, and T. Frenkiel,
 Adv. Magn. Res. {\bf 11}, 47 (1983).

\bibitem{dec2} A.~J. Shaka, J. Keeler, T. Frenkiel, and R. Freeman, J.
Magn.\ Reson.\ {\bf 52}, 335 (1983).

\bibitem{dec3} U. Eggenberger, P. Schmidt, M.
Sattler, S.~J. Glaser, and C. Griesinger, J. Magn.\ Reson.\ {\bf 100},
604 (1992).

\bibitem{geodesics}
N. Khaneja, B. Heitmann, A. Sp\"orl, H. Yuan, T. Schulte-Herbr\"uggen, S. J. Glaser,
Phys. Rev. A {\bf 75}, 012322 (2007).

\bibitem{time-opt-tril}
 N. Khaneja, S.~J. Glaser, R. Brockett,
Phys. Rev. A {\bf 65}, 032301 (2002).

\end{thebibliography}
 \end{document}